\documentstyle[12pt]{article}
\begin{document}
\begin{center}
{\large \bf 
Anomalous Behaviors in Fractional Fokker-Planck Equation 
}\\

\vspace*{.5in}

\normalsize 
Kyungsik Kim$^{*}$\\

\vspace*{.2in}

{\em Department of Physics, Pukyong National University,\\
Pusan 608-737, Korea}\\

\vspace*{.2in}
\hfill\\

Y. S. Kong \\

\vspace*{.2in}

{ \em School of Ocean Engineering, Pukyong National University,\\
Pusan 608-737, Korea}\\

\vspace*{.5in}
\hfill\\
\end{center}
\baselineskip 24pt
\vskip 5mm 
\begin{center}
{\bf Abstract}
\end{center} 

\noindent
We introduce a fractional Fokker-Planck equation with a temporal power-law dependence on 
the drift force fields. For this case, the moments of the tracer from the force-force
correlation in terms of the time-dependent drift force fields are discussed
analytically. The long-time asymptotic behavior of the second moment is determined by 
the scaling exponent $\xi$ imposed by the drift force fields.
In the special case of the space scaling value $\nu=1$ and the time scaling value 
$\tau=1$, our result can be classified 
according to the temporal scaling of the mean second moment
of the tracer for large $t$: $< \overline{x^2(t)} >$ $\propto$ $t$ with $\xi=\frac{1}{4}$ 
for normal diffusion, and $< \overline{x^2(t)} >$ $\propto$ $t^{\eta}$ with $\eta>1$ 
and $\xi>\frac{1}{4}$ for superdiffusion.   

$^{*}$E-mail: kskim@pknu.ac.kr                                             
\newpage

\noindent

Recently, a lot of interest has been concentrated  on nonstationary random
processes in  the behavior of disordered systems $[1-3]$.
The anomalous feature of diffusive stochastic processes
is known to be qualitatively different from the standard behavior characteristic of
the regular system. Especially, the anomalous diffusive behavior in the disordered system has
mainly been argued from the random walk, a useful theory of classical
stochastic processes in the past; in fact, the normal diffusion of tracers
is well known to be a result of random walkers characterized by  the variance
$<x^2 (t)>$ $\propto$ $ t$.  The random walk theory
with the nearest-neighboring transition probability has been numerically discussed
by using the Green function method.  The anomalous transport process
of the random walk theory
has been widely extended to a continuous-time random walk
theory, which is specified by the distributions of both the pausing time and
the transition probability, which are dependent on the length between steps.
It provides a dynamic framework for a precise connection
between fractional diffusion equations and fractal walks $[4,5]$
while the transport phenomena for the motions of tracers
have largely been extended to reaction kinetics$[6,7]$ and strange kinetics $[8]$.

It is well known that the second moment for the superdiffusive proess
is proportional to $t^{\alpha}$ with $\alpha>1$
for superdiffusion and $\alpha<1$ for subdiffusion
and to $log^{\beta} t$ for strong diffusion.
In the past, the best example of
the superdiffusive random proess was a classical paper by Richardson $[9]$ in which
the second moment for a relative separation $l$ at time $t$, $<l^2 >$ 
$\propto$ $t^3 $,
was shown to depend on the motion of two particle in a turbulent fluid.
Matheron and de Marsily $[10]$ showed the anomalous behavior of the longitudinal dispersion,
$<x^2 (t) >$ $\propto$ $t^{3/2}$, in a stratified random velocity field.
Among other recent examples,
we also can mention anomalous kinetics in chaotic dynamics due to flights and trapping
$[8]$, layered velocity fields$[11]$,
anomalous diffusion of amphiphilic molecules $[12]$, anomalous diffusion
in a two-dimensional motion of rotating flow $[13]$, etc. Examples for subdiffusive
stochastic processes are charge transport in amorphous semiconductors $[14,15]$, 
NMR diffusometry on percolation structures $[16]$, and the dynamics of a bead in polymers $[17]$.
Furthermore, the stochastic process for anomalous diffusion on a fractal structure and
L$\acute{e}$vy fleights has been the subject of several reports in the literature $[18 - 20]$.
The typical models using L$\acute{e}$vy fleights include bulk-mediated surface diffusion $[21]$, 
transport inmicelle systems or heterogeneous rocks $[22]$, single molecule spectroscopy
$[23]$, and special reaction dynamics$[24]$. Among the many frameworks 
for characterizing anomalous diffusion are fractional Brownian motion $[25]$, 
fractional diffusion equations $[26,27]$, and generalized Langevin and Fokker-Planck equations $[28-31]$.

The motivation for this paper was to apply the analytical method of a field-theoretical 
space-time renormalization group to investigate the anomalous behavior of the 
fractional diffusion equation. The moments for the probability density will be found
so that the properties lead to anomalous asymptotics of the diffusion process.  
The purpose of this paper is to present an anomalous transport process for a test particle 
described by a fractional Fokker-Planck equation in the presence of a temporal power-law 
dependence for the force-force correlation of the drift force fields. The resulting formulae for
the diffusive behavior of the tracer in the Fourier-Laplace region are discussed 
analytically, and the anomalously long-time behavior of the moments can be determined 
by using the scaling exponent $\xi$ imposed by drift force fields. 

First, the fractional diffusion equation for the probability density $P(r,t)$ on the 
$d$-dimensional fractal structure can be expressed in terms of
\begin{equation}
\frac{\partial^{\beta}}{\partial t^{\beta}}  P(r,t)  = 
\frac{1}{r^{d-1}} 
\nabla_{\alpha}  D(r)r^{d-1}  \nabla_{\alpha}  P(r,t),
\label{eq:a11}
\end{equation}
where $\nabla_{\alpha} = \frac{\partial^{\alpha}}{\partial
(-r)^{\alpha} }$ and $D(r)$ is the anomalous diffusivity and depends only on the position. 
The normalization condition for the probability density is given by 
\begin{equation}
\int^{\infty}_0 Dr^{d-1} P(r,t) dr =  1.
\label{eq:b22}
\end{equation}
If $\alpha = \beta = 1$, the form of Eq. $(1)$ transforms 
to a spherically symmetric diffusion equation in $d$-dimensional Euclidean 
space, showing that this equation has analytical solutions for a 
diffusive behavior characteristic of fractal systems. 
Let us now introduce the anomalous diffusivity given by
\begin{equation}
D(r)  = Dr^{\theta},
\label{eq:c33}
\end{equation}
where $D$ is the diffusion constant and $\theta$ is the scaling exponent and can be 
determined from the fractal structure. The feature of this diffusivity 
describes the statistical property of diffusion on fractals, and 
even though it extends to a number of $\theta$ and $\alpha $ 
we here consider the special case of $ \theta = 
\alpha $ and $ 0 < \alpha < 1 $ in order to find the moments.
After multiplying Eq. $(1)$ by $r^{\alpha}$ and  $r^{2 \alpha}$, respectively, 
and integrating, we obtain the asymptotic result 
\begin{equation}
< r^2 (t) >  \sim  t^{2\beta / \alpha}.
\label{eq:d44}
\end{equation}
This represents the behavior for a superdiffusive particle for the scaling exponent
relation $2\beta > \alpha$ and a subdiffusive particle for $0 < 2\beta < \alpha$.
From Eq. $(4)$, the value of the scaling exponent, $2\beta = \alpha$ is 
equal to one for normal diffusion.

From now, we introduce the equation of motion in the presence of
a spatiotemporal power-law dependence of both the drift force and an
external harmonic potential.
Since we take into account an overdamped particle of mass $m$ moving in a viscous
medium. The equation of motion can be expressed in terms of 
\begin{equation}
m \ddot{\vec{r}} = f(\vec v,t)-V'(\vec r)+\Gamma(t), 
\label{eq:a1}
\end{equation} 
where $f(\vec v, t)=-\gamma(t)\vec v$ is a time-dependent drift force field, 
and a harmonic force is given by $F(r)= -V^{\prime}(\vec r)$.
The term $\Gamma(t)$ is a fluctuating force with 
$<\Gamma(t)>=0$ and $< \Gamma(t_1) \Gamma(t_2)>= 2D\delta(t_1 - t_2 )$. From Eq. $(5)$ 
we can find the dispersive behavior of a test particle, and that mathematical techique 
leads us to a more general result.

In the three-dimensional case of an external harmonic potential $V(r)={\frac{1}{2}}$ $a r^2$,
the formal solution of Eq. $(5)$ is  
\begin{equation}
\vec r (t)  =  \vec r (0) Y(t) + Y(t) \int_{0}^{t} \frac{\Gamma(t^{'})}{Y(t^{'})}dt^{'},
\label{eq:c3}
\end{equation}
where  $Y(t) = exp[-\int_{0}^{t}\hat{\gamma}(t^{'})dt^{'}]$ and $\hat{\gamma}(t^{'})
= \gamma(t^{'})+ a$. Then, one can show that a direct calculation of
the variance gives
\begin{equation}
\sigma^{2} (t)=De^{-2at} Y^{2} (t) \int_{0}^{t} {e^{-2at'} \over Y^{2} (t^{'}) }dt^{'} 
\label{eq:d4}
\end{equation}

From Refs. $32$ and $33$, the quantity $\sigma^{2} (t)$ is easily 
compared with that of other results.
If $\gamma(t)$ in terms of the time-dependent drift force field is given by a
general form, we can discuss extentively the anomalous diffusive behavior of the tracer. 
Therefore, the Fokker-Planck equation associated with Eq. $(5)$ is 
\begin{equation}
\frac{\partial}{\partial t} P(\vec r, t)= - \bigtriangledown \cdot [\gamma(t)\vec r 
+ \vec F(\vec r)]P(\vec r, t) 
+D {\bigtriangledown}^{2} P(\vec r, t).
\label{eq:e5}
\end{equation}

Especially, we will focus on two fractional Fokker-Planck equations:
\begin{eqnarray}
\frac{\partial}{\partial t} P(\vec{r},t)&=& - \bigtriangledown  \cdot [ \vec{F} (\vec{r},t)
\frac{\partial^{1-\tau}}{\partial t^{1-\tau}}P(\vec{r},t)] \nonumber \\
&+& D {\bigtriangledown}^{2} \frac{\partial^{1-\tau}}{\partial t^{1-\tau}} P(\vec{r}, t)
\label{eq:f6}
\end{eqnarray}
and 
\begin{equation}
\frac{\partial}{\partial t} P(\vec{r},t)= - \bigtriangledown  \cdot [ \vec{F} (\vec{r},t)P(\vec{r},t)]
+ D {\bigtriangledown}^{\nu} P(\vec{r}, t)
\label{eq:f66}
\end{equation}
with $\frac{1}{2} <\tau \leq 1$ and $1 <\nu \leq 2$, where $\tau$ and 
$\nu$ are, respectively, the time and the space scaling exponents. 
In Eq. $(10)$, the Fourier-Laplace transform of the probability density function is described as
\begin{equation}
P(\vec{k}, \omega) = \widetilde{\it L}\widetilde{\it F}[P(\vec r, t) \Theta(t)],
\label{eq:g7}
\end{equation}
where $\Theta(t)$ is a step function, and $\widetilde{\it F}$ and $\widetilde{\it L}$ are,
respectively, the Fourier and the Laplace transform operators. In the Fourier-Laplace transform domain, 
Eqs. $(9)$ and $(10)$ become
\begin{eqnarray}
& &[-i \omega +D \vert \vec k \vert^{2} {(-i\omega)}^{1-\tau} ]{P(\vec k, \omega)} = \nonumber \\
& &1- { {i \vec k} {(-i\omega)}^{1-\tau} \over (2\pi)^{d+1}} \int d \vec k^{'} d \omega'
\vec F(\vec k - \vec k^{'}, \omega - \omega^{'}) P(\vec k^{'}, \omega^{'})  \nonumber \\
\label{eq:i9}
\end{eqnarray}
and
\begin{eqnarray}
&&[-i \omega +D \vert \vec k \vert^{\nu} ]{P(\vec k, \omega)} = \nonumber \\
&&1- { {i \vec k} \over (2\pi)^{d+1}} \int d \vec k^{'} d \omega'
\vec F(\vec k - \vec k^{'}, \omega - \omega^{'}) P(\vec k^{'}, \omega^{'}).  \nonumber \\
\label{eq:i999}
\end{eqnarray}
Since $\Lambda$$(\vec k, \omega)$ and $\Xi$$(\vec k, \omega)$ in the two cases 
where Eqs. $(2)$ and $(3)$ are, respectively, defined by
\begin{equation}
\Lambda (\vec k, \omega) = \frac{{(-i\omega)}^{\tau-1} }{ {(-i \omega)}^{\tau} +D \vert \vec k \vert^{2}},
   \Xi(\vec k, \omega) = \frac{ 1 }{ {(-i \omega)}^{\tau} +D \vert \vec k \vert^{2}}  \nonumber \\
\label{eq:i99}
\end{equation}
and 
\begin{equation}
\Lambda (\vec k, \omega) = \frac{1 }{ {-i \omega} +D \vert \vec k \vert^{\nu}},
   \Xi(\vec k, \omega) = \frac{ 1 }{ {-i \omega} +D \vert \vec k \vert^{\nu}},
\label{eq:i99}
\end{equation}
Eqs. $(12)$ and $(13)$ can be put into a unitary form:
\begin{eqnarray}
P(\vec k, \omega) &=& \Lambda (\vec k, \omega)
-\frac{i } {(2\pi)^{d+1} }  \nonumber \\
&\times& \int d \vec k^{'}
d \omega^{'} G(\vec k, \vec k - \vec k^{'}, \omega - \omega^{'})P(\vec k^{'}, \omega^{'}),  \nonumber \\
\label{eq:l11}
\end{eqnarray}
where
\begin{equation}
G(\vec k, \vec k - \vec k^{'}, \omega - \omega^{'} ) = \Xi(\vec k, \omega)\vec k \cdot 
F(\vec k - \vec k^{'}, \omega - \omega^{'}).
\label{eq:j10}
\end{equation}

For the special case of a tracer in the presence of a time-dependent force,
the force-force correlation can be described as

\begin{equation}
<\vec{F} (\vec{r}, t ) \vec{F} ({\vec r}^{'}, t^{'} )> = b^2 {(t t^{'})}^{\xi} 
\delta(\vec r -{\vec r}^{'} ) \delta(t - {t}^{'} ) 
\label{eq:o14}
\end{equation}
and
\begin{eqnarray}
&&<\vec{F} ({\vec k} , \omega ) \vec{F} ({\vec k}^{'} , \omega^{'} )>
={2 \pi b^2} \Gamma ( 2 \beta + 1) \nonumber \\
&\times& {( -i \omega - i \omega^{'} )}^{-( 2 \xi + 1 )}
 \delta ( \vec{k}_{1} + \vec{k}_{2} ).
\label{eq:o14}
\end{eqnarray}
Hence, from the inverse Fourier-Laplace transform, the $n$-$th$ moment of the tracer
can be given by 
\begin{equation}
\overline{x^n (t)}  = \widetilde{\it L}^{-1} [ i^{n} \frac{\partial^{n}} {\partial {k_x}^{n}} P(\vec k, \omega)
]_{\vec k = 0},
\label{eq:m121}
\end{equation}
where $\widetilde{\it L}^{-1}$ is the inverse Laplace-transform operator.
In order to investigate the relaxation dynamics of the probability density 
toward the stationary solution,
we now restrict our attention to the special class with the space scaling value $\xi=1$ 
and the time scaling value $\tau=2$. 
We find the statistical behavior of a tracer from the moments of the probability density.
From our fractional Fokker-Planck equation via Eqs. $(12)-(19)$, the mean second moment
for large $t$ is 
\begin{equation}
< \overline{x^2(t)} > \simeq 2Dt + {\frac{ 4 {\pi}^{1/2} b^{2} }{ (2\xi +1) D^{1/2}}} 
t^{ {2 \xi} + { \frac{1}{2} }} .
\label{eq:q15}
\end{equation}
\\
From Eq. $(21)$, $< \overline{x^2(t)} >$ $\propto$ $t$ with $\xi=\frac{1}{4}$ 
for normal diffusion and $< \overline{x^2(t)} >$ $\propto$ $t^{\eta}$ with $\eta>1$ and $\xi>\frac{1}{4}$ 
for superdiffusion.   

In conclusion, from a fractional Fokker-Planck equation, the tracer dipersion for
an anomalous transport process has been discussed in the presence of a temporal power-law dependence 
of the drift and an external harmonic potential.
The mean moments of the tracer were analyzed for the subdiffusive or superdiffusive
behavior for several values of the exponent $\xi$ in the temporal power-law dependence of the drift.
We think that for arbitrary time and space scaling exponents, it may not be easy
to discuss the diffusive behaviors of the fractional Fokker-Planck equation.
We also will present the results for non-Fickian diffusion for arbitrary time and space scaling exponents
in other journals.
We expect the detailed description of
the anomalous behavior to be used to study extensions of numerical stratified models
and fractal lattice models.

\vskip 5mm
\begin{center}
{\bf ACKNOWLEDGMENT}
\end{center}
\hfill\\

This work was supported by grant No. 2000-2-133300-001-3 from the Basic Research 
Program of the Korea Science and Engineering Foundation.

\vspace {3mm}
%

\end{document}